\documentclass[conference]{IEEEtran}

\usepackage{wrapfig} 
\usepackage{hyperref}
\usepackage{amsmath}
\usepackage{algorithm}
\usepackage{algorithmic}
\usepackage{multirow}   
\usepackage{xcolor}
\usepackage{graphicx}
\usepackage{booktabs}
\usepackage{pgfplots}
\usepackage[utf8]{inputenc}
\usepackage[most]{tcolorbox}
\usepackage{balance}

\pgfplotsset{compat=1.16}

\title{Can Large Language Models Generate Observability-Aware Code?
}
\author{
\IEEEauthorblockN{
Yongliang Tao$^{1}$,
Hongyu Zhang$^{1}$,
Pengfei Gao$^{2}$,
Minghua Ma$^{2}$,
Zhiyu Fan$^{2}$,
Yu Kang$^{2}$
}

\vspace{0.3em}

\IEEEauthorblockN{
Jue Zhang$^{2}$,
Si Qin$^{2}$,
Liqun Li$^{2}$,
Qingwei Lin$^{2}$,
Saravan Rajmohan$^{2}$
}

\vspace{0.5em}

\IEEEauthorblockA{
$^{1}$Chongqing University, Chongqing, China
}

\IEEEauthorblockA{
$^{2}$Microsoft
}

}

\begin{document}
\maketitle

\begin{abstract}

Recent advances in coding agents have enabled the generation of increasingly complex software systems. While existing evaluations primarily focus on functional correctness, production systems must expose failure evidence to support observability. 
In this paper, we present a systematic study of observability in agent-generated systems. We examine whether agents can reconstruct source-level diagnostic semantics by restoring observability artifacts in 10 open-source and 8 industrial repositories. We also evaluate whether these artifacts translate into effective fault signals at runtime through 200 generated microservice systems deployed on Kubernetes with 13 injected faults. 
Our results reveal a consistent gap between diagnostic semantics at the source level and fault signals (i.e., explicit, fault-specific evidence) at runtime. At the source level, agents partially recover observability artifacts but struggle to capture key diagnostic semantics. At runtime, generated systems expose fault signals for only a small fraction of failures (up to 13.99\%), despite the presence of logging, suggesting that the generated observability artifacts may lack the failure-specific semantics needed to effectively expose faults. We further introduce an observability-oriented skill, which can serve as a guidance to improve both diagnostic semantics and fault-signal exposure, but the gains remain limited, indicating that the gap is not easily addressed. 
More broadly, our findings suggest that current evaluations focusing primarily on functional correctness may overlook observability as an important dimension of practical software quality.
\end{abstract}




\section{Introduction}
Recent advances in coding agents, such as Copilot~\cite{microsoft2026copilot}, 
Cursor~\cite{cursor2026}, and Claude Code~\cite{anthropic2025claudecode}, 
have made it increasingly feasible to generate complete software systems. 
These systems may compile, deploy, and run successfully under normal workloads. However, software that is runnable is not necessarily software that is operable. When failures occur, developers need sufficient information to understand what happened. This shift from runnability to operability brings failure-time observability into focus.
Beyond generating runnable code, agents should also generate observability artifacts (i.e., logs, traces, and metrics) that expose fault-indicating runtime evidence.

In human-written software systems, even though developers may have a substantial understanding of the system, unexpected production failures still occur, making observability a fundamental need for debugging.

Coding agents further increase developers' need for  observability. While they can generate large amounts of code quickly, developers may not inspect or reason about all of the generated code in detail. As a result, developers may not fully understand the runtime behavior of the generated system. This creates \textbf{knowledge debt}: developers may need to maintain generated systems whose runtime behavior they do not fully understand.

Knowledge debt makes observability more difficult to improve after systems are deployed. 
When developers lack comprehensive understanding of the runtime behavior of agent-generated systems, they may lack the semantic context needed to determine what diagnostic semantics (i.e., failure-relevant context captured by observability artifacts) should be encoded. This is because observability design requires reasoning about potential failure modes and identifying relevant runtime states, which depends on a deep understanding of system behavior. Consequently, post-hoc instrumentation may fail to capture a failure-relevant context and, therefore, fail to expose the corresponding fault signal during failures.

We therefore suggest that observability should be treated as a generation-time expectation for agent-generated code. As coding agents implement functionality, they should simultaneously generate \textbf{observability artifacts},  so that diagnostic semantics are encoded during code generation and the fault signal can be exposed when failures occur.

This reframes observability from a post-hoc remediation activity into an integral part of code generation. Accordingly, this paper asks \textbf{whether current coding agents generate code that is not only runnable but also observable under failure?}

To answer this question, we connect source-level diagnostic semantics (i.e., failure-relevant context captured by observability artifacts, such as error codes or request IDs) with the fault signal (i.e., explicit, fault-specific evidence available during failure-time execution, such as a  service crash observed in logs). Source-level analysis evaluates whether generated observability artifacts encode meaningful diagnostic semantics, while runtime analysis evaluates whether the generated systems expose the corresponding fault signal under injected failures. Together, these two perspectives enable a comprehensive evaluation of observability in agent-generated code.

Accordingly, we structure our study around the following research questions:

\noindent\textbf{RQ1 (Static Observability Restoration):} Can coding agents generate observability artifacts consistent with human-written observability artifacts?

\textit{Method.} We conduct a controlled source-level restoration study on 10 open-source and 8 industrial systems with mature human-written observability artifacts, covering 1,223 instances in total. We remove existing observability artifacts while preserving the original business logic, ask coding agents to restore them using the remaining repository context, and evaluate both the placement of generated observability artifacts and the diagnostic semantics they capture.

\textit{{Findings.}} Coding agents recover observability only partially. They are substantially better at identifying where observability artifacts should be placed than what diagnostic semantics should be recorded, revealing a persistent semantic observability gap even when the functional business logic is already available.

\noindent\textbf{RQ2 (Observability in Agent-Generated Microservice Systems):} To what extent do end-to-end agent-generated microservice systems expose fault signals under realistic failure scenarios?

\textit{Method.} 
We generate 200 microservice systems from high-level specifications using coding agents with SOTA LLMs, deploy them in Kubernetes, and inject 13 representative production faults, resulting in 1{,}615 failure instances. We collect runtime logs during each fault window and measure \textbf{Fault Signals Rate (FSR)}, the fraction of injected faults that produce explicit fault signals in logs.

\textit{{Findings.}} Agent-generated systems produce fault signals for only a small fraction of failures, with FSR ranging from 4.95\% to 13.99\% across models. The limitation is not the absence of logging, but the systematic absence of explicit fault-indicating semantics and diagnostic context in the generated logs.




\noindent\textbf{RQ3 (Observability-Oriented Guidance):}
Can lightweight observability-oriented guidance (Skill) improve both static diagnostic semantics and runtime fault signals exposure?

\textit{Method.} We derive a lightweight observability-oriented skill from approximately 200 real-world commits, summarizing diagnosis-oriented practices (e.g., instrumentation placement and diagnostic data capture) as structured generation guidance. We then re-evaluate the enhanced agents under both the source-level restoration (RQ1) and runtime fault-injection (RQ2) settings.

\textit{Findings.} Observability-oriented guidance yields only marginal improvements in static diagnostic semantics, ranging from 0.004 to 0.012 across models. For runtime fault exposure, while FSR increases modestly for GPT-5.5 (+8.67 pp), the gains for Claude Opus 4.8 (+0.99 pp) and Gemini 3.5 Flash (+2.54 pp) remain minimal, leaving the overall FSR relatively low (up to 16.53\%). These limited gains suggest that lightweight generation-time guidance may partially mitigate, but is unlikely to fully resolve, the semantic observability gap, indicating substantial room for more effective methods.




The major contributions of the paper are as follows:

  \begin{itemize}
  \item \textbf{A multi-level framework for evaluating agent-generated observability.}
  We introduce a multi-level evaluation framework that examines agent-generated code from two complementary perspectives. Specifically, the framework evaluates, at the static level, the ability of coding agents to restore observability artifacts in terms of their placement and the diagnostic semantics they capture; at the dynamic level, it assesses whether the generated observability implementations can effectively expose fault signals under real system deployment and fault injection scenarios.

  \item \textbf{Empirical evidence of a semantic observability gap.}
Across 10 open-source and 8 industrial systems, agents recover human-written observability only partially: they are better at placing observability artifacts than recovering diagnostic semantics. In 200 generated microservice systems and 1{,}615 injected failure instances, logs are widely generated but often uninformative, failing to expose meaningful fault signals under failure conditions. As a result, only a small fraction of failures can be associated with actionable runtime evidence, indicating a substantial gap in failure-time observability.

  \item \textbf{Implications for observability-aware code generation.}
Our results show that current coding agents exhibit insufficient observability in generated code. Lightweight observability-oriented guidance improves fault signal exposure but still leaves a substantial gap, suggesting that future coding agents and benchmarks should treat observability as a first-class objective in code generation and evaluation.
  \end{itemize}

\section{Background and Motivation}

\begin{figure*}[t]
   \centering
   \includegraphics[width=0.9\textwidth]{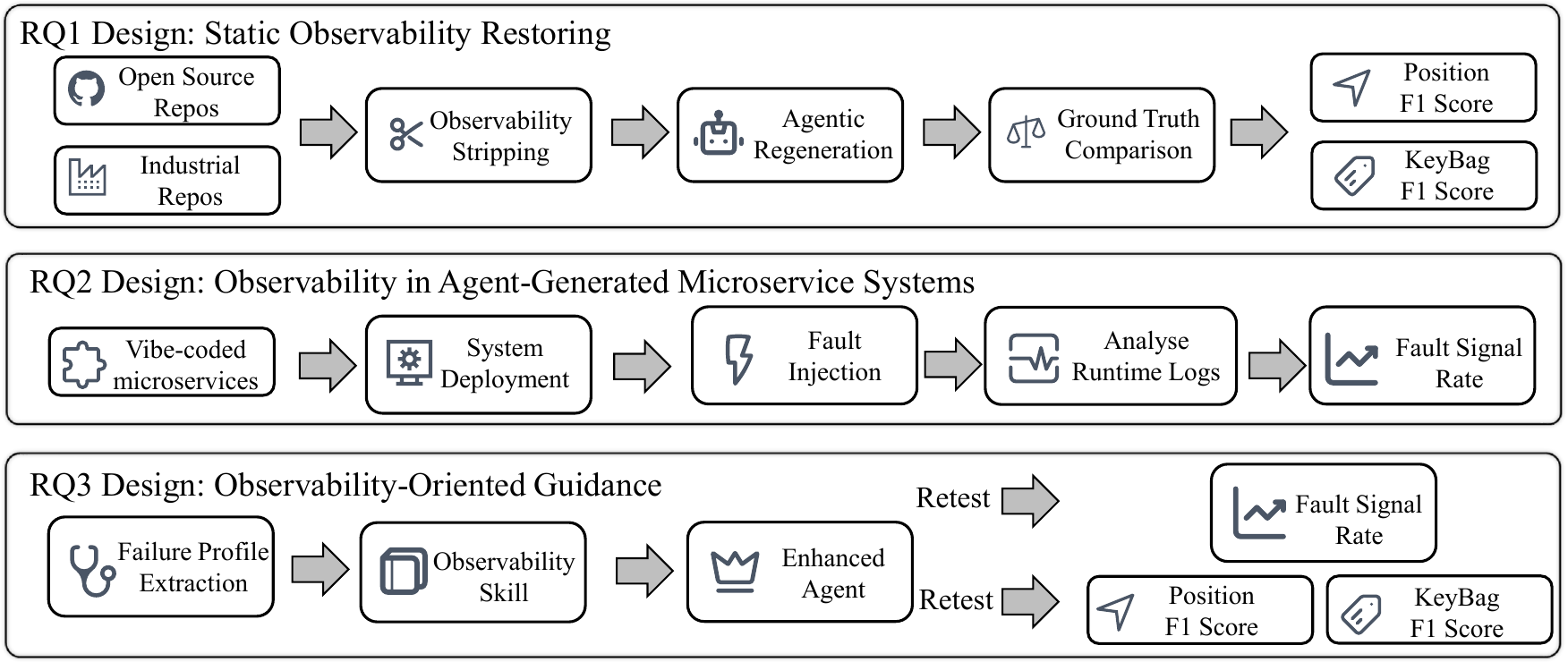}
   \caption{Overview of Study}
   \label{logextract}
   \vspace{-6pt}
\end{figure*}

\subsection{Agent-Generated Code Creates Knowledge Debt}
The increasing adoption of coding agents changes how developers acquire operational understanding of software systems. Traditionally, implementation and understanding are tightly coupled: by constructing software themselves, developers gradually build mental models of execution flow, dependency interactions, and potential failure modes.

Coding agents weaken this coupling and further increase developers’ reliance on observability. While they can generate large amounts of code quickly, developers may not inspect or reason about all of the generated code in detail. As a result, they may not fully understand the runtime behavior of the generated system. We refer to this gap as {knowledge debt}: developers may need to maintain systems whose runtime behavior they do not fully understand, making diagnosing failures in production significantly more difficult.

Observability has long been essential for production diagnosis. In the era of coding agents, however, runtime evidence increasingly compensates for reduced authorship-based understanding. Consequently, the key question is not only whether coding agents can generate functionally correct software, but whether they can also generate observability that supports effective failure-time diagnosis.

\subsection{From Diagnostic Semantics to Fault Signals}

We study observability across two levels. 
 At the \emph{source-code level}, an observability artifact encodes \textbf{diagnostic semantics}: the failure-relevant context it is intended to capture, such as dependency states, error codes, request  identifiers, state transitions, and variables that explain exceptional behavior. 
 At the \emph{runtime level}, observability should produce a \textbf{fault signal}: explicit, fault-specific evidence emitted during failure-time execution, such as a dependency becoming unavailable, a request  exceeding its timeout, or an invalid state transition. 
 In this study, observability refers to the ability of generated code to encode diagnostic semantics in source code and expose fault signals at runtime.

 These two levels are connected but not equivalent. 
 Diagnostic semantics are a precondition for fault signals: instrumentation that does not encode failure-relevant context cannot expose such context at runtime. 
 However, source-level semantics alone do not guarantee runtime evidence, because the relevant execution path must be exercised, the generated functionality must behave as intended, and the evidence must remain  visible in logs. 
 Thus, source-level analysis captures whether agents express meaningful diagnostic intent, while runtime analysis captures whether generated systems expose that intent under real failures.

Evaluating coding agents across both levels is essential for understanding whether they can generate observability code that goes beyond functional correctness and exposes failure-relevant evidence in production-like settings.

\section{Study Design}
\subsection{Overview}

Our study investigates the following central 
question:
\textit{to what extent can current coding agents generate code with effective observability?} 
To answer this question, we design a three-stage evaluation pipeline that progressively examines observability from isolated capability analysis to realistic runtime behavior and guided improvement, as illustrated in Figure~\ref{logextract}.

\begin{figure*}[t]
   \centering
   \includegraphics[width=0.9\textwidth]{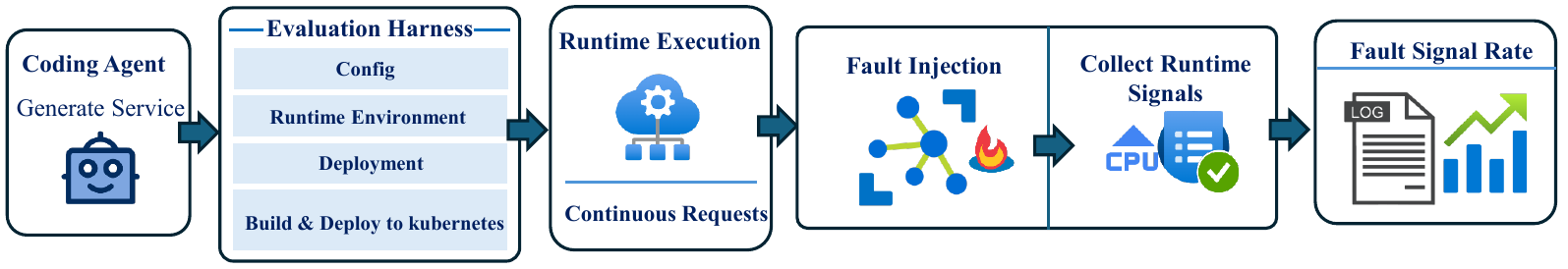}
   \caption{Evaluation of Observability in Agent-Generated Microservice Systems}
   \label{rq2_agent}
\end{figure*}

\textbf{RQ1 (Static Observability Restoration).}
To isolate the underlying capability of coding agents in capturing diagnostic semantics without the confounding burden of business logic generation, we conduct a controlled source-level reconstruction study. We curate 1,223 instances from 10 open-source and 8 industrial repositories, remove their human-written observability artifacts, and task coding agents with reconstructing them. This controlled, repository-aware setting enables us to evaluate the agents' capability to recover diagnosis-oriented observability artifacts independently of business logic synthesis. We assess reconstruction quality using \emph{Position F1} (placement accuracy) and \emph{KeyBag F1} (semantic overlap of essential diagnostic variables).

\textbf{RQ2 (Observability in Agent-Generated Microservice Systems).}
Building upon the capability analysis in RQ1, we investigate whether the observed source-level capability limitation is consistently or not reflected in realistic runtime behavior. We generate 200 microservice systems from high-level specifications, deploy them in a Kubernetes environment, and inject 13 representative production faults, resulting in 1,615 failure instances. We analyze runtime logs using the \emph{Fault Signals Rate (FSR)}, which measures the proportion of injected faults that expose explicit fault signals.

\textbf{RQ3 (Observability-Oriented Guidance).}
Finally, we investigate whether structured diagnosis-oriented guidance can improve the observability capability of coding agents. We derive a lightweight observability skill from real-world observability-related commits and apply them during system generation. The enhanced agents are then re-evaluated under the same source-level restoration and runtime fault-injection settings. 

Together, these three research questions form a unified multi-level evaluation framework. RQ1 isolates the capability of coding agents to model diagnostic semantics in a controlled setting. RQ2 evaluates whether this capability limitation is consistently reflected in the runtime fault-signal exposure of end-to-end generated systems. RQ3 investigates how lightweight diagnosis-oriented guidance can further improve runtime fault-signal exposure, providing insights into the development of future observability-aware coding agents.

\subsection{RQ1 Design: Static Observability Restoration}
\textbf{Goal.} RQ1 asks whether agents can restore human-written source-level observability at the code level, after the original observability statements are removed.
\subsubsection{Dataset}

We construct a dataset from both open-source and industrial repositories, including 10 GitHub repositories~\cite{otel_demo,microservices_demo_google,deathstarbench,eshop,nestjs,robusta,microservices_demo_other,train_ticket,strapi,vector} (441 instances) and 8 industrial datasets (782 instances), resulting in 1,223 instances in total.

The selected GitHub repositories are widely used microservice projects, generally with more than 1k stars, and are actively maintained by their communities. These projects exhibit sustained demand for observability improvements, where logging statements, telemetry instrumentation, and related observability artifacts are continuously introduced and refined to support operational diagnosis.

The selected industrial repositories are maintained over extended periods (typically longer than one year), deployed in real production environments, and require observability improvements driven by incidents. In these settings, missing observability signals are often identified and introduced during incident diagnosis and operational maintenance.


\subsubsection{Procedure} 

\begin{figure}[t]
   \centering
   \includegraphics[width=0.5\textwidth]{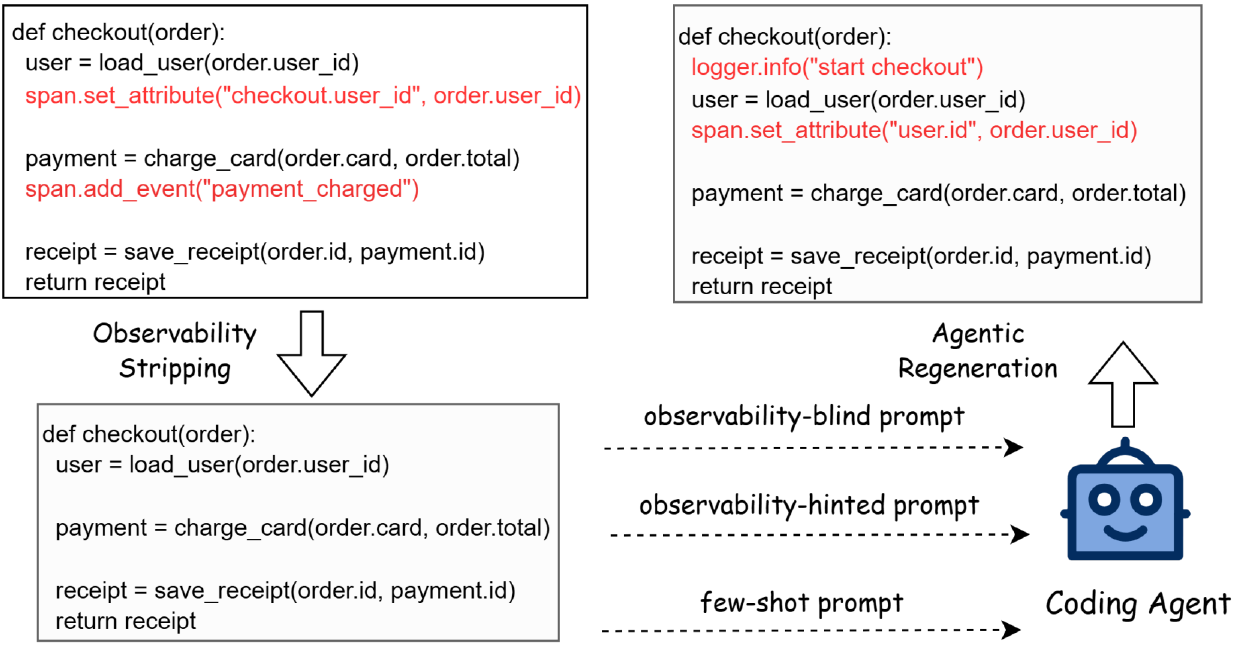}
   \caption{Example of Static Observability  Restoration }
   \label{codeagentExample}
\end{figure}
Instances of static observability restoration are extracted using language-specific parsers and
retained if they contain at least one observability construct, including logging statements, tracing operations, metric updates,
span attributes, or span events. We exclude trivial functions,
test files, generated code, and build artifacts.
For each instance, we remove all observability statements
(e.g., logging and tracing) while preserving business logic
and control-flow structures.
The stripped implementation serves as the agent input,
whereas the original implementation is retained as the
ground truth (Figure~\ref{codeagentExample}).

Importantly, the agent operates over the \textbf{entire repository context}, rather than a single function in isolation. 
This setting distinguishes coding agents from standalone LLMs, allowing them to exploit repository-level conventions such as logging styles, tracing schemas, and telemetry naming patterns.

\subsubsection{Prompting Strategies}
We evaluate three prompting configurations:

\textbf{Observability-blind:} no explicit instruction on observability; to assess whether coding agents can proactively introduce missing observability artifacts without explicit guidance, reflecting their inherent observability awareness.

\textbf{Observability-hinted:} explicit instruction to improve observability; to evaluate how explicit guidance influences the quality and completeness of generated observability artifacts.

\textbf{Few-shot:} same-file examples of observability instances; to evaluate whether coding agents can learn and adopt repository-specific observability practices from in-context example.






\subsubsection{Functional Scope}

RQ1 is a static source-level restoration study.
Generated code is not executed during this evaluation, and we
do not assess whether the regenerated implementation preserves
the original functional behavior.
Instead, the evaluation focuses exclusively on whether coding
agents recover human-written observability artifacts, including
their placement and diagnostic semantics.

\subsubsection{Evaluation Metrics}

Because both the original and reconstructed implementations are available, we can evaluate observability reconstruction along two complementary aspects: \emph{where} artifacts are placed and \emph{what} they capture.

\paragraph{Position F1 (Where to instrument)}

Position F1 measures whether generated observability appears at the same execution
regions as the ground truth. We first align generated and ground-truth functions
using non-observability statements as anchors. These anchors partition each
function into buckets: before the first matched statement, between consecutive
matched statements, and after the last matched statement.

Each bucket is represented by a binary indicator denoting whether it contains
any observability statement. A bucket is counted as a true positive if both the
ground truth and generated code contain observability in that aligned bucket, as
a false positive if only the generated code does, and as a false negative if
only the ground truth does.

Let $TP_{\mathrm{pos}}$, $FP_{\mathrm{pos}}$, and $FN_{\mathrm{pos}}$ denote
these bucket-level counts. We compute:

\[
\textstyle
\mathrm{Position\ P} =
\frac{TP_{\mathrm{pos}}}{TP_{\mathrm{pos}} + FP_{\mathrm{pos}}},
\
\mathrm{Position\ R} =
\frac{TP_{\mathrm{pos}}}{TP_{\mathrm{pos}} + FN_{\mathrm{pos}}}.
\]

\[
\mathrm{Position \ F1} =
\frac{2 \cdot \mathrm{Position\ P} \cdot \mathrm{Position\ R}}
{\mathrm{Position\ P} + \mathrm{Position\ R}}
\]

This metric evaluates whether observability is restored at the correct workflow
locations, independently of the exact diagnostic content.

\paragraph{KeyBag F1 (What to capture)}

Position correctness alone does not guarantee that the restored observability
captures the same diagnostic concepts. We therefore compute a token-level
content metric over aligned buckets where both the ground truth and generated
code contain observability.

For each such bucket, we extract a bag of normalized lexical tokens from the
observability statements. The extractor considers string literals, identifiers,
attribute names, and keyword-argument names, then normalizes them by lowercasing,
splitting on common separators such as dots, underscores, hyphens, slashes, and
camel-case boundaries, and removing observability-framework stop words such as
\texttt{logger}, \texttt{span}, and \texttt{event}.

Let $T^{gt}_b$ and $T^{gen}_b$ be the resulting token sets for a comparable
bucket $b$. We aggregate token-level true positives, false positives, and false
negatives across all comparable buckets:

\[
TP_{\mathrm{key}} = \sum_b |T^{gt}_b \cap T^{gen}_b|,
\]

\[
FP_{\mathrm{key}} = \sum_b |T^{gen}_b \setminus T^{gt}_b|,
\
FN_{\mathrm{key}} = \sum_b |T^{gt}_b \setminus T^{gen}_b|.
\]

We then compute:

\[
\textstyle
\mathrm{KeyBag\ P} =
\frac{TP_{\mathrm{key}}}{TP_{\mathrm{key}} + FP_{\mathrm{key}}},
\;
\mathrm{KeyBag\ R}  =
\frac{TP_{\mathrm{key}}}{TP_{\mathrm{key}} + FN_{\mathrm{key}}}
\]

\[
\mathrm{KeyBag\ F1} =
\frac{2 \cdot \mathrm{KeyBag\ P} \cdot \mathrm{KeyBag\ R}}
{\mathrm{KeyBag\ P} + \mathrm{KeyBag\ R}}
\]

KeyBag F1 is therefore a lightweight lexical proxy for diagnostic-content
agreement. It rewards generated observability that mentions the same operational
concepts as the ground truth, while Position F1 separately accounts for missing
or spurious instrumentation locations.

\subsection{RQ2 Design: Observability in Agent-Generated Microservice Systems
}
\textbf{Goal.} As the observability of systems generated by coding agents in practical settings remains largely unclear, RQ2 measures the Fault Signals Rate in agent-generated systems to assess whether generated systems can produce fault signals under faults.


\subsubsection{Dataset}

We collect 200 representative architecture specifications for cloud-native microservice applications, by abstracting architectures from real-world open-source and industrial systems, covering a diverse range of domains, including e-commerce, social networking, IoT telemetry, finance, healthcare, content management, SaaS platforms, travel, logistics, gaming, and observability/DevOps systems. Each architecture specification is provided to the coding agent as the implementation prompt, allowing the agent to automatically generate a complete deployable microservice system for subsequent runtime evaluation under fault injection.

\subsubsection{Procedure}

Figure~\ref{rq2_agent} presents the end-to-end runtime fault-injection evaluation pipeline. For each architecture specification, the coding agent generates a deployable microservice implementation, which is subsequently executed within a standardized evaluation harness. The harness provides a unified runtime environment, deployment configuration, workload generation, and fault-injection infrastructure, ensuring that all generated systems are evaluated under the same settings.

After deployment and successful health verification, each service is continuously exercised by the workload generator while representative production faults are injected using Chaos Mesh. Runtime logs are collected throughout each fault window and analyzed offline to determine whether the generated observability exposes diagnostic evidence corresponding to the injected fault.

\subsubsection{Fault Injection}
\begin{table*}[t]
\centering
\caption{Evaluated fault primitives with injection and log evidence.}
\label{tab:faults_combined}
\small
\begin{tabular}{lllll}
\hline
\textbf{ID} & \textbf{Fault} & \textbf{Chaos Type} & \textbf{Injected Failure} & \textbf{Summary of Fault-specific Log Evidence} \\
\hline
F01 & pod-kill      & PodChaos     & Kill service pod 
    & restart/startup evidence or restartCount increase \\
F02 & network-delay & NetworkChaos & Service egress latency 
    & timeout/latency/deadline evidence \\
F03 & upstream-fail & HTTPChaos    & Upstream returns 503 
    & upstream/HTTP 503 \\
F04 & upstream-slow & NetworkChaos & Upstream latency 
    & upstream timeout/slow/deadline evidence \\
F05 & db-down       & PodChaos     & Postgres pod kill 
    & postgres/connection refused/FATAL evidence \\
F06 & db-slow       & NetworkChaos & Postgres latency 
    & db timeout/query latency/deadline evidence \\
F07 & cache-down    & PodChaos     & Redis cache pod kill 
    & redis/connection reset/refused evidence \\
F08 & cache-slow    & NetworkChaos & Redis cache latency 
    & redis timeout/slow/cache latency evidence \\
F09 & queue-down    & PodChaos     & Redis stream pod kill 
    & stream/queue/redis connection failure evidence \\
F10 & queue-slow    & NetworkChaos & Redis stream latency 
    & queue/stream timeout or slow publish/consume evidence \\
F11 & cpu-stress    & StressChaos  & CPU pressure 
    & CPU/load/slow processing/timeout evidence \\
F12 & net-corrupt   & NetworkChaos & Packet corruption 
    & reset, broken pipe, network error evidence \\
F13 & time-skew     & TimeChaos    & Clock offset 
    & clock/time skew/token expiry/timestamp evidence \\
\hline
\end{tabular}
\end{table*}
To evaluate runtime observability under realistic operating conditions, we inject representative production faults using Chaos Mesh~\cite{mesh} in a Kubernetes cluster. As summarized in Table~\ref{tab:faults_combined}, our benchmark covers 13 fault primitives spanning service failures, dependency failures, network anomalies, resource contention, and time-related faults. These fault types simulate common failure scenarios encountered in cloud-native microservice systems, including service crashes, upstream failures, database and cache outages, queue disruptions, CPU contention, network corruption, and clock skew.
Across the 200 generated systems, these configurations produce a total of 1,615 executable fault instances. 

\subsubsection{Evaluation Metrics}
After fault injection, we allow sufficient time  for fault propagation and collect runtime 
logs from all affected services. We focus on logs as the primary evaluation signal because 
they are the \textbf{most widely used} and consistently available observability artifacts produced by current coding agents, whereas traces and metrics are often incomplete or entirely absent in generated systems. As a result, logs provide the most reliable basis for assessing failure visibility. If even logs \textbf{fail to expose fault signals} effectively, this suggests that the observability of current coding agents is fundamentally limited.

\textbf{Conservative Notion of Observability.} 
It is important to clarify that our evaluation adopts a conservative notion of observability. 
We do not require logs to fully diagnose root causes or support complete Root Cause Analysis~\cite{sui2023logkg,icseticket,FaultLocalization,10.1145/3627703.3629553,rca_icse,rcafse} (RCA). 
Instead, a log is considered a fault signal only if it explicitly encodes failure-related semantics, such as dependency failures, timeout conditions, protocol errors, or other fault-specific indicators that can be directly attributed to the injected fault. 
Generic information such as request logs or standard status codes is not sufficient unless it explicitly reflects abnormal system behavior. 
This design intentionally avoids relying on post-hoc inference from runtime symptoms and ensures that detected signals correspond to explicit and actionable fault semantics.

For each fault type, we define a set of fault-specific log evidence (see Table~\ref{tab:faults_combined}) derived from the expected 
manifestations of the injected failure. Examples include timeout-related messages for network 
delays, connection-refused errors for service disruptions, and out-of-memory indicators for 
resource exhaustion. These signatures serve as a proxy for the observability of failures in logs. 
A fault instance is considered \emph{observable} if at least one corresponding fault signal
appears in the collected logs after fault injection.

Based on this definition, we compute the Fault Signals Rate (FSR) as $FSR = \frac{N_{\text{Signal}}}{N_{\text{fault}}}$, where $N_{\text{Signal}}$ denotes the number of fault instances that produce observable fault-related signals, and $N_{\text{fault}}$ denotes the total number of injected fault instances

Unlike traditional observability evaluations that focus on the quantity of generated logs, 
metrics, or traces, this evaluation measures whether generated observability artifacts 
actually expose runtime failures. A low FSR indicates that failures remain largely 
silent, preventing subsequent diagnosis regardless of the volume of generated observability artifacts.

This evaluation enables us to quantify the practical effectiveness of observability in 
LLM-generated systems and to examine whether the observability artifacts introduced by coding 
agents are sufficient to surface failures in realistic deployment environments.

\subsection{RQ3 Design: Observability-Oriented Guidance}

\textbf{Goal.}
RQ3 investigates whether the static and runtime observability gaps identified in RQ1 and RQ2 can be mitigated through targeted, diagnosis-oriented observability guidance.

\textbf{Dataset and System.}
To ensure comparability with prior findings, we reuse the same set of application generation tasks and fault-injection scenarios from RQ1 and RQ2. By reusing the same workload, we can directly measure whether the proposed mitigation reduces the previously observed gaps.


\textbf{Procedure.} We construct an observability skill by analyzing approximately 200 observability-related code commits collected from an internal failure repository. These commits were created by experienced engineers to diagnose and remediate production failures, providing practical examples of effective observability artifacts. They capture recurring diagnosis and remediation patterns, including dependency failures, timeouts, retries, and exception propagation. We abstract these recurring patterns into a set of lightweight, diagnosis-oriented principles.

We then evaluate a \emph{skill-guided agent}, where the observability skill is provided as structured generation guidance, and compare it against a \emph{baseline agent} without such guidance. 
Both agents are tasked with generating or instrumenting services under the same workload and fault conditions. 
All other factors (model, prompts, and environment) are held constant to isolate the effect of the skill.

\textbf{Skill Content.}
Due to space constraints, we present a high-level abstraction of the observability skill. 
The skill consists of diagnosis-oriented principles that guide the agent to:

\begin{itemize}
    \item Identify service boundaries, responsibilities, dependencies, and operational contexts.
    \item Instrument dependency calls and control-flow boundaries where failures or latency may occur.
    \item Capture key diagnostic signals, including correlation IDs, status codes, latency, retry counts, timeout values, and exception causes.
    \item Follow or establish consistent observability conventions for service, operation, dependency, and request context.
    \item Ensure that instrumentation remains lightweight, avoids high-cardinality or sensitive data, and supports efficient diagnosis under failure scenarios.
\end{itemize}

\subsection{Implementation and Reproducibility}

We evaluate three closed-source frontier models: GPT-5.5~\cite{openai}, Claude Opus 4.8~\cite{anthropic}, and Gemini 3.5 Flash~\cite{google}. 
All model calls are made through a unified agent interface based on GitHub Copilot SDK~\cite{copilot_sdk}, using the model identifiers exposed by the provider at experiment time. 
For generated-service experiments, all models share identical task descriptions, prompt templates, repository workspaces, and interaction protocols. 
The agent operates with repository access and automatic tool-permission approval, and execution traces are recorded for auditing.
\begin{table*}[t]
\caption{Performance comparison across prompting strategies}
\centering
\small
\begin{tabular}{llcccccc}
\toprule
Prompt & Model & Pos. P & Pos. R & Pos. F1 
& KeyBag P & KeyBag R & KeyBag F1 \\
\midrule

\multirow{3}{*}{observability-blind prompt}
& GPT-5.5        & 0.544 & 0.606 & 0.551 & 0.331 & 0.458 & 0.357 \\
& Claude Opus 4.8 & 0.554 & 0.708 & 0.580 & 0.270 & 0.412 & 0.294 \\
& Gemini 3.5 Flash &  0.277  &  0.333  &  0.278   &  0.233  &  0.262   &  0.237   \\

\midrule

\multirow{3}{*}{observability-hinted prompt}
& GPT-5.5        & 0.442 & 0.695 & 0.505 & 0.204 & 0.488 & 0.260 \\
& Claude Opus 4.8 & 0.426 & 0.842 & 0.528 & 0.243 & 0.476 & 0.291 \\
& Gemini 3.5 Flash  &  0.269   &  0.438   &  0.307   &  0.176  &  0.288   &  0.197 \\

\midrule

\multirow{3}{*}{few-shot examples prompt}
& GPT-5.5        & 0.530 & 0.774 & 0.594 & 0.334 & 0.571 & 0.383 \\
& Claude Opus 4.8 & 0.568 & 0.816 & 0.633 & 0.371 & 0.518 & 0.398 \\
& Gemini 3.5 Flash  & 0.409   &  0.578  &  0.448   &  0.349   &  0.446   &  0.367  \\

\bottomrule
\end{tabular}
\label{tab:prompting_results}
\end{table*}





\section{Study Results}
\subsection{RQ1 Results: Static Observability Restoration}
To answer RQ1, we systematically investigate how coding agents perform under the three prompting strategies introduced earlier. We obtain the following findings.


\textbf{Finding 1: Implicit observability awareness exists, but artifact quality remains limited.}
As shown in Table~\ref{tab:prompting_results}, coding agents demonstrate a degree of implicit observability awareness even under the \textit{observability-blind} prompt, as they still recover a non-trivial portion of observability-related content. For instance, GPT-5.5 achieves a KeyBag recall of 0.458, while Claude Opus 4.8 reaches positional recall of 0.708, indicating that these models can recognize when observability artifacts are needed even without explicit guidance.

\begin{figure}[t]
   \centering
   \includegraphics[width=0.45\textwidth]{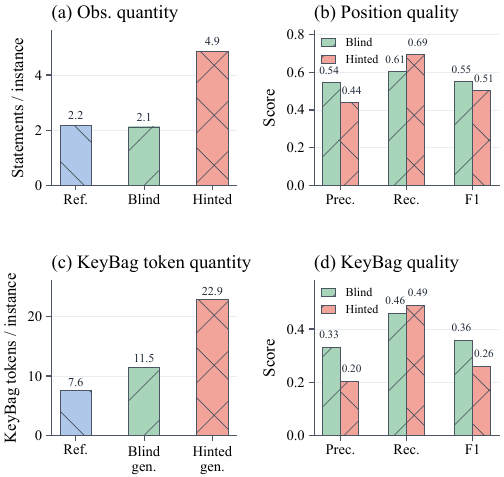}
   \vspace{-10pt}
   \caption{Quantity over Quality under observability-hinted prompt}
   \vspace{-5pt}
   \label{quality_quantity}
\end{figure}

However, this awareness does not translate into high-quality outputs. Under blind prompting, the KeyBag F1 scores remain low across all models (0.357 for GPT-5.5, 0.294 for Claude Opus 4.8, and 0.237 for Gemini 3.5 Flash), suggesting that the generated artifacts are frequently incomplete, noisy, or misaligned with the expected key elements.

\textbf{Finding 2:  Quantity over Quality under Explicit Instructions.}
As shown in Fig.~\ref{quality_quantity}, explicit instructions (i.e., observability-hinted prompts) primarily encourage increased generation rather than improved quality, compared to implicit instructions (i.e., observability-blind prompts). 

From the quantity perspective, Fig.~\ref{quality_quantity}(a) and (c) show that explicit instructions lead to substantially more generated statements (4.9 vs. 2.1) and KeyBag tokens (22.9 vs. 11.5), indicating clear over-generation. We measure quantity as the average number of generated observability statements and the number of corresponding KeyBag tokens per instance.

However, this increased quantity does not translate into better quality. 
In Fig.~\ref{quality_quantity}(b) and (d), precision drops notably while recall improves only slightly. 
Specifically, KeyBag recall increases (0.488 vs. 0.458, +0.030), but precision decreases sharply (0.20 vs. 0.33, -0.13), resulting in a lower F1 score (0.26 vs. 0.36, -0.10). 
A similar pattern is observed for position quality, where F1 also declines (0.51 vs. 0.55, -0.05).

This reveals a potential quantity–quality trade-off: explicit instructions push the model toward over-generation, improving coverage at the expense of correctness. 
Such over-generation not only reduces output quality but also introduces unnecessary computational and post-processing overhead, posing an additional burden on downstream systems.

\textbf{Finding 3: Contextual Few-Shot Improves Both Content and Placement Quality.}
As shown in Table~\ref{tab:prompting_results}, Contextual few-shot prompting improves both the content quality and placement accuracy of generated observability artifacts. Compared to implicit instructions, few-shot prompting increases KeyBag F1 from 0.357 to 0.383 (+0.025) and Position F1 from 0.551 to 0.594 (+0.043). 

These improvements are primarily driven by substantial gains in recall (KeyBag: +0.114; Position: +0.168), while maintaining comparable precision. This suggests that few-shot examples help the model better identify relevant observability signals without introducing excessive noise.

We attribute this improvement to the contextual grounding provided by in-context examples, which enables the model to better align with repository-specific observability practices and reduce ambiguity during generation.

\textbf{Finding 4: Placement is Easier than Content Generation.}
As shown in Table~\ref{tab:prompting_results}, across all prompting strategies, Position F1 consistently exceeds KeyBag F1, suggesting that agents more reliably determine where to introduce observability than what observability artifacts to generate. For example, under implicit prompting, Position F1 (0.551) is significantly higher than KeyBag F1 (0.357), with similar gaps under explicit prompting (0.505 vs. 0.260) and few-shot prompting (0.594 vs. 0.383). This reveals a gap between structural understanding and content-level precision, suggesting that coding agents can capture structural signals more reliably than semantic requirements of observability content.

\newenvironment{summarybox1}{
  \vskip 5pt
  \begin{center}
  \colorbox{gray!25}{
    \parbox{0.95\linewidth}{
      \normalsize 
      \textbf{Summary.} Observability is not a simple instruction-following capability but a context-dependent generation problem. Coding agents show partial awareness of observability, but generated artifacts often lack precision and fail to record key diagnostic
semantics, reflected in consistently low KeyBag F1 scores. This raises a key question: whether such source-level limitations persist in realistic settings, particularly in the ability of generated systems to expose fault signals during runtime?
    }%
  }
  \vskip 2pt
  \end{center}
}{}

\begin{summarybox1}
\end{summarybox1}

\begin{table}[t]
\centering
\small
\setlength{\tabcolsep}{5pt}
\renewcommand{\arraystretch}{1.2}
\caption{RQ2 runtime observability under fault injection. }
\label{tab:rq2_runtime}
\begin{tabular}{lccc}
\toprule
\textbf{Model} & \textbf{Runnable Services} & \textbf{FSR} & \textbf{Subset FSR} \\
\toprule
GPT-5.5 & 151 / 200 & 4.95\% & 6.56\% \\
Claude Opus 4.8 & 154 / 200 & 6.32\% & 8.28\% \\
Gemini 3.5 Flash & 136 / 200 & 13.99\% & 20.62\% \\
\bottomrule
\end{tabular}
\end{table}

\subsection{RQ2 Results: Observability in Agent-Generated Microservice Systems}

To answer RQ2, we evaluate runtime observability using the Fault Signal Rate (FSR), defined as the fraction of intended fault instances for which the generated system emits fault-specific log evidence during the fault window. Each model is evaluated on 1,615 intended fault instances. We report both the overall FSR and the Subset FSR (computed only over successfully executed services).


\textbf{Finding 1: Most injected faults remain unobservable.}
As shown in Table~\ref{tab:rq2_runtime}, all models exhibit consistently low FSR, ranging from 4.95\% to 13.99\%. 
Even when restricting the evaluation to successfully executed services, the best-performing model achieves only \textbf{20.62\%}, indicating that the majority of injected faults are not reflected as explicit fault signals in runtime logs.

This result does not imply that generated systems lack logging. 
Many services emit runtime logs, including request processing events and status codes. 
However, these logs often do not encode fault-specific semantics that can be attributed to particular failure types. 
Thus, the limitation is not the absence of logging, but the absence of explicit failure semantics in logs.
\begin{figure}[t]
   \centering
   \includegraphics[width=0.5\textwidth]{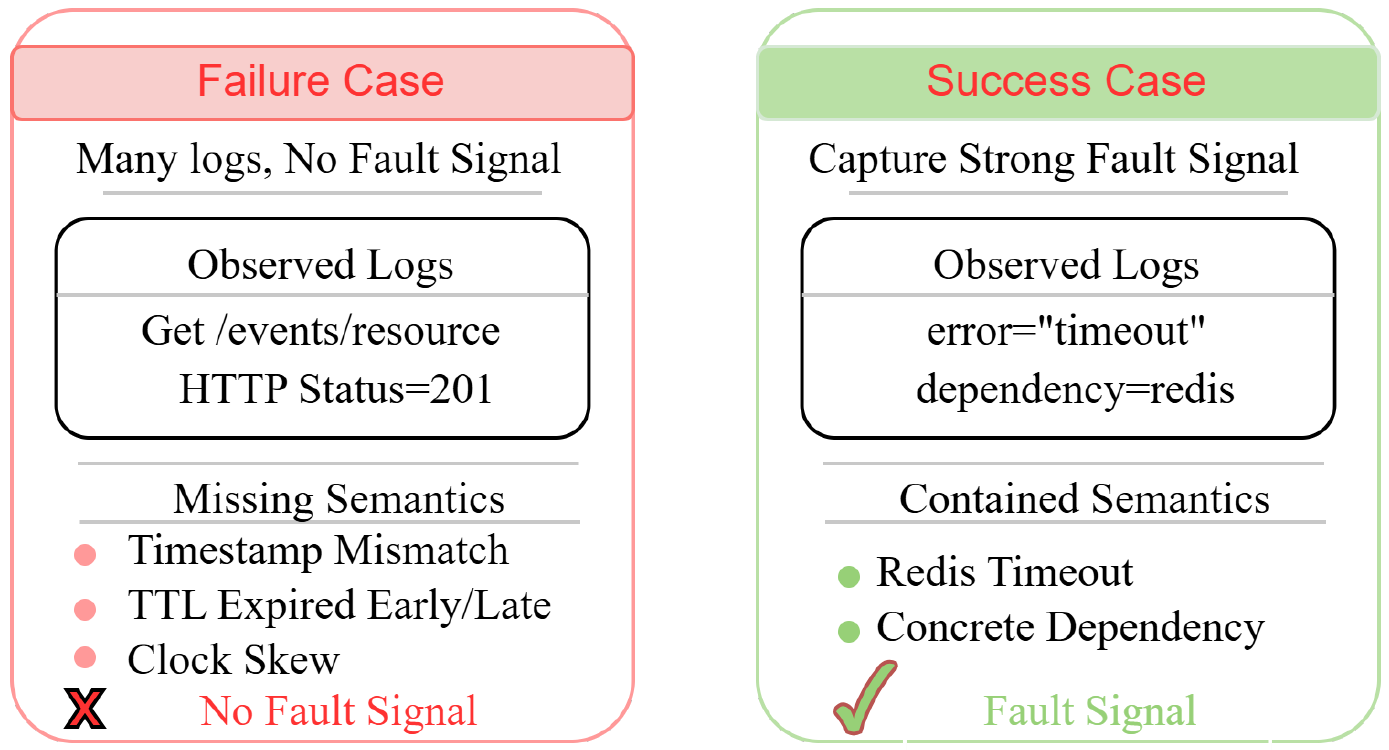}
   \vspace{-10pt}
   \caption{Representative success and failure cases under the same evaluation. }
   \vspace{-5pt}
   \label{two_case}
\end{figure}

\begin{table}[t]
\centering
\small
\setlength{\tabcolsep}{5pt}
\renewcommand{\arraystretch}{1.2}
\caption{Per-Fault FSR Under Fault Injection. 
}
\label{tab:rq2_fault_difficulty}
\begin{tabular}{lccc}
\toprule
\textbf{Fault} & \textbf{Caught / Total} & \textbf{FSR} & \textbf{Difficulty} \\
\midrule
F03 (upstream fail) & 14 / 51  & 27.45\% & Easy \\
F01 (pod kill)      & 122 / 600 & 20.33\% & Easy \\
F07 (cache down)    & 55 / 321  & 17.13\% & Easy \\
F08 (cache slow)    & 46 / 321  & 14.33\% & Easy \\
\midrule
F10 (queue slow)    & 18 / 198  & 9.09\%  & Medium \\
F02 (network delay) & 53 / 600  & 8.83\%  & Medium \\
F05 (db down)       & 32 / 420  & 7.62\%  & Medium \\
F09 (queue down)    & 15 / 198  & 7.58\%  & Medium \\
\midrule
F06 (db slow)       & 25 / 420  & 5.95\%  & Hard \\
F04 (upstream slow) & 3 / 51    & 5.88\%  & Hard \\
F12 (net corrupt)   & 10 / 555  & 1.80\%  & Hard \\
F11 (cpu stress)    & 8 / 555   & 1.44\%  & Hard \\
F13 (time skew)     & 7 / 555   & 1.26\%  & Hard \\
\bottomrule
\end{tabular}
\end{table}

\noindent
\textbf{Finding 2: Observable faults are primarily those with explicit error manifestations.}
FSR varies significantly across fault types (Table~\ref{tab:rq2_fault_difficulty}). 
Faults that naturally produce explicit errors—such as upstream failures (27.45\%), pod crashes (20.33\%), and cache outages (17.13\%)—achieve substantially higher detection rates, as they directly manifest as connection failures, error responses, or service unavailability. 
\textbf{As illustrated in our success cases (Figure~\ref{two_case}), when such explicit errors occur, the generated logs successfully capture failure-related semantics (e.g., affected dependencies, concrete timeout messages), allowing the faults to be directly exposed as fault signals.}

In contrast, faults with weaker or more implicit effects—such as time skew (1.26\%), CPU stress (1.44\%), and network corruption (1.80\%)—exhibit consistently low coverage. 
These faults do not inherently generate explicit error messages and can only be observed if the system proactively records additional runtime context.

\noindent
\textbf{Finding 3: Generated logs lack diagnostic semantics.}
Although generated systems frequently emit runtime logs, these logs rarely include the information required to distinguish between different failure types. 
In particular, generated logs seldom explicitly encode affected dependencies, failure modes, or abnormal system conditions. 
{This is clearly demonstrated in our failure cases (Figure~\ref{two_case}): the system emits multiple runtime logs (e.g., request events, normal HTTP 201 status codes), but completely lacks explicit failure semantics (e.g., timestamp mismatches or clock drift). The logs exist in abundance, but they are semantically hollow regarding the actual fault.}

As a result, even when faults occur and logs are present, the available information is often insufficient to determine what kind of failure has happened. 
This observation is consistent with the pattern in Finding 2: faults that inherently produce explicit errors can still be detected, whereas faults requiring contextual interpretation remain unobservable because the generated logs fail to express the necessary failure-relevant information.

\newenvironment{summarybox2}{
  \vskip 5pt
  \begin{center}
  \colorbox{gray!25}{
    \parbox{0.95\linewidth}{
      \normalsize 
      \textbf{Summary.} The observability of microservices generated by current coding agents appears limited, with at most 13.99\% of injected faults being exposed. This limitation does not seem to arise from the absence of logs, but is associated with the inability of logs to capture fault-relevant signals. This finding is consistent with the low KeyBag F1 scores observed in RQ1. 
    }%
  }
  \vskip 2pt
  \end{center}
}{}

\begin{summarybox2}
\end{summarybox2}

\subsection{RQ3 Results: Observability-Oriented Guidance}
\begin{table}[t]
\caption{Effect of observability skill on diagnostic semantics and fault signals.}
\centering
\scriptsize
\setlength{\tabcolsep}{4pt}
\renewcommand{\arraystretch}{0.92}
\begin{tabular}{llccc}
\toprule
\multicolumn{1}{c}{Model} & Setting 
& \multicolumn{1}{c}{Pos F1} 
& \multicolumn{1}{c}{KeyBag F1} 
& \multicolumn{1}{c}{FSR} \\
\midrule
\multicolumn{1}{c}{GPT-5.5} 
& w/o skill & 0.594 & 0.383 & 4.95\% \\
& w/ skill  & 0.597 & 0.395 & 13.62\% \\
& Gain      & $\Delta$0.003 & $\Delta$0.012 & +8.67 pp \\
\midrule
\multicolumn{1}{c}{Claude Opus 4.8} 
& w/o skill & 0.633 & 0.398 & 6.32\% \\
& w/ skill  & 0.649 & 0.407 & 7.31\% \\
& Gain      & $\Delta$0.015 & $\Delta$0.009 & +0.99 pp \\
\midrule
\multicolumn{1}{c}{Gemini 3.5 Flash} 
& w/o skill & 0.448 & 0.367 & 13.99\% \\
& w/ skill  & 0.452 & 0.371 & 16.53\% \\
& Gain      & $\Delta$0.004 & $\Delta$0.004 & +2.54 pp \\
\bottomrule
\end{tabular}
\label{tab:skill_static_dynamic}
\end{table}
RQ1 shows the limitations in generating diagnostic semantics, and RQ2 shows that generated systems exhibit low FSR under failures. To examine whether such limitations can be easily mitigated, we evaluate whether or not an explicit observability skill can improve system observability under the same setting.
The results are shown in Table~\ref{tab:skill_static_dynamic}.

\textbf{Finding 1: Improvements in fault signals.}
Observability skill improves FSR across all evaluated coding agents. GPT-5.5 increases from 4.95\% to 13.62\% (+8.67 pp), Claude Opus 4.8 from 6.32\% to 7.31\% (+0.99 pp), and Gemini 3.5 Flash from 13.99\% to 16.53\% (+2.54 pp). However, the overall FSR remains relatively low.


\textbf{Finding 2:  Limited improvements in diagnostic semantics.}
In addition to FSR, both Pos F1 and KeyBag F1 show small but consistent improvements across models (see Table~\ref{tab:skill_static_dynamic}), indicating that observability skill can partially enhance source-level diagnostic semantics.

\newenvironment{summarybox3}{
  \vskip 5pt
  \begin{center}
  \colorbox{gray!25}{
    \parbox{0.95\linewidth}{
      \normalsize 
      \textbf{Summary.} Overall, observability skill improves FSR, Pos F1, and KeyBag F1 to a limited extent. While these results suggest that skill-based guidance can enhance observability, the remaining gap indicates substantial room for future methods to further improve observability-aware code generation. 
    }%
  }
  \vskip 2pt
  \end{center}
}{}

\begin{summarybox3}
\end{summarybox3}

\section{Discussion and Threats to Validity}
\subsection{Discussion}
This study examines the observability of agent-generated microservice systems from two complementary perspectives. At the source-code level, observability restoration evaluates whether coding agents can recover the diagnostic semantics embedded in human-written observability code. The results show that agents are able to restore only part of these observability artifacts, indicating limited capability in encoding fault-relevant diagnostic semantics. At the system level, observability evaluation measures whether the generated systems expose actionable fault signals during failures. Despite producing runnable systems with abundant logging output, the generated systems expose fault signals for only a small fraction of injected faults. Taken together, these findings suggest that current coding agents remain insufficient in generating observability that effectively supports failure diagnosis.

To investigate whether this limitation can be mitigated during generation, we introduce lightweight observability-oriented skills. Although these skills consistently improve FSR, Position F1, and KeyBag F1, the overall gains remain modest. This suggests that simply augmenting prompts with observability-related guidance is insufficient to substantially improve diagnostic quality. Instead, generating effective observability appears to require deeper reasoning about failure propagation, runtime states, and diagnostic semantics, which current prompt-level interventions do not fully provide.

Overall, our findings indicate that observability should be considered an independent capability of coding agents rather than a by-product of functional code generation. Future research may require integrating failure-aware reasoning or runtime feedback into the generation process, instead of relying solely on prompt engineering.

One possible explanation is that observability is rarely represented as an independent objective in the training data. In real-world software repositories, logging and other observability-related code are typically developed together with functional code and are gradually refined during software evolution. Consequently, coding agents may learn where observability components are usually inserted, allowing them to reproduce common logging structures, while failing to infer what diagnostic information should be recorded for specific failures.

Another possible explanation is that current coding agents primarily learn from static source code rather than runtime system behavior. Effective observability depends on understanding how failures manifest during execution and which runtime states provide actionable diagnostic evidence. Without \textbf{deployment }or execution feedback during training, agents may lack sufficient knowledge to reason about failure propagation, resulting in logs that execute correctly but rarely expose informative fault signals.

\subsection{Threats to Validity}
We have identified the following major threats to validity.

\textbf{Oracle-based FSR may miss valid fault signals.}
The definition of Fault Signals Rate (FSR) relies on fault-specific oracles to identify valid fault signals. Although the oracle is constructed based on failure-relevant patterns, it may not capture all possible forms of valid signals. In particular, some logs may contain partial or indirect information that does not satisfy the oracle criteria and is therefore counted as \textit{no signal}. However, since the same oracle is consistently applied across all evaluated settings, and our goal is to measure the presence of actionable fault signals rather than arbitrary logging outputs, the relative comparisons and overall trends remain stable.

\textbf{Human-written observability does not represent an optimal upper bound.}
Our study derives diagnostic semantics (KeyBags) from human-written code. Many of these observability-related signals are introduced or refined after faults have occurred, reflecting practical debugging and operational processes. While such code captures realistic engineering practice, it does not guarantee complete or optimal observability. In some cases, human-written systems may still omit fault-relevant signals or include redundant logging. Therefore, our evaluation reflects alignment with practical engineering standards rather than an ideal ground truth. Importantly, coding agents still fail to reproduce these practical diagnostic patterns even under simplified settings such as observability reconstruction, suggesting that the observed limitation is unlikely to be caused by the choice of reference.

\textbf{Randomness of LLM-based generation may affect FSR.}
Coding agents based on large language models exhibit inherent randomness in generation. Variations in decoding may affect the placement and content of generated logs, which in turn can influence the observed FSR. To mitigate this effect, we follow previous works~\cite{liuloglm} and repeat each experiment five times, reporting the average results. We further note that LLM hallucination~\cite{hull,hull2} may introduce additional variability in generated logs; however, the consistency of trends across repeated runs suggests that randomness does not materially affect our conclusions.


\section{RELATED WORK}
\textbf{Evaluation of Coding Agents: }
The rapid advancement of large language models has led to a wide range of benchmarks~\cite{LiveCodebench,FeatBench,mbpp,deveval,CoderEval} for evaluating coding agents. Early studies primarily focused on function-level code generation, with datasets such as HumanEval~\cite{chen2021humaneval} and Terminal-Bench~\cite{merrill2026terminalbench} assessing functional correctness through unit test pass rates. These benchmarks emphasize whether generated programs execute correctly, but largely abstract away system-level complexity.
More recent repository-level benchmarks such as SWE-Bench~\cite{jimenez2024swebench} extend evaluation to realistic software engineering tasks, where agents are required to resolve issues in full GitHub repositories. This line of work better captures engineering complexity, but evaluation remains largely centered on task completion or patch correctness.
As a result, they provide limited insight into runtime behavior under failure conditions, particularly whether generated systems expose fault signals for debugging. This reveals a gap between functional correctness-oriented evaluation and observability in real-world deployment scenarios.

\textbf{Software Observability and Logging Practices: }
Software observability, particularly logging, is a fundamental practice for understanding and debugging complex systems. Prior work shows that effective logs should provide rich contextual information, such as variable states, dependency interactions, and error propagation paths, rather than simple execution traces~\cite{fu2014logging}. 
As systems evolve toward distributed and microservice architectures, observability becomes increasingly critical, since failures often arise from cross-service interactions~\cite{obs1}. Moreover, logging is typically not designed upfront but evolves iteratively as developers encounter failures and augment instrumentation with additional diagnostic context. 



\textbf{Log Analysis and Instrumentation: }
Automated log analysis has evolved from heuristic methods to recent deep learning and large language models, demonstrating strong semantic reasoning in Log Parsing~\cite{he2017drain,jiang2024logparsing,logparse3}, Log Anomaly Detection~\cite{logllm,logprompt,raglog,li2020_swisslog,logcluster,evlog,lou2010mining,deeplog,logsentiment,plelog}, and  Root Cause Analysis~\cite{eadro,tracelog,rootcause}. 
Despite these advances, log analysis and instrumentation remain largely decoupled. Existing instrumentation approaches focus on surface-level patterns and rarely ensure that generated logs encode sufficient failure-relevant semantics for effective runtime diagnosis. As a result, whether automated coding agents can generate observability artifacts with rich diagnostic semantics remains an open question.

\section{Conclusion}
In this paper, we present a systematic study of observability in code generated by coding agents. 
The lack of diagnostic semantics observed in source-level observability restoration, together with the low FSR under fault injection, suggests that current coding agents generate code with \textbf{limited observability}. 
We further introduce a lightweight observability-oriented guidance, which improves both diagnostic semantics and fault signals exposure. While these results demonstrate the potential of observability-oriented guidance, they also show that the improvements remain {limited}, and reveal substantial room for future research toward enabling coding agents to consistently generate observability-aware code. We call for future research to focus on enabling coding agents to generate observability-aware code that can consistently produce fault signals reflecting underlying failures. 



\bibliographystyle{IEEEtran.bst}
\bibliography{sample-base}

\end{document}